\documentclass[pdflatex,sn-nature]{sn-jnl}

\usepackage{graphicx}
\usepackage{amsmath,amssymb}
\usepackage{booktabs}
\usepackage{multirow}
\usepackage{xcolor}
\usepackage{textcomp}
\usepackage{siunitx}
\DeclareSIUnit\angstrom{\text{\AA}}
\usepackage[version=4]{mhchem}

\begin{document}

\title[Nonlinear optical activation in \ce{Mo2TiC2T_x} MXene]
{\texorpdfstring{\ce{Mo2TiC2T_x}}{Mo2TiC2Tx} MXene Saturable Absorption for Neural Networks on a Chip}

\author[1]{\fnm{Shadad} \sur{Watad}}
\author[1]{\fnm{Aviad} \sur{Katiyi}}
\author[2]{\fnm{Bar} \sur{Favelukis}}
\author[3]{\fnm{Muhammad Sharif} \sur{Uddin}}
\author[4]{\fnm{Anupma} \sur{Thakur}}
\author[2]{\fnm{Maxim} \sur{Sokol}}
\author[3,5]{\fnm{Babak} \sur{Anasori}}
\author*[1]{\fnm{Alina} \sur{Karabchevsky}}

\affil*[1]{\orgdiv{School of Electrical and Computer Engineering}, \orgname{Ben-Gurion University of the Negev}, \orgaddress{\city{Beer-Sheva}, \postcode{8410501}, \country{Israel}}}

\affil[2]{\orgdiv{Department of Materials Science and Engineering},\orgname{Tel Aviv University},\orgaddress{\city{Tel Aviv}, \postcode{6997801},\country{Israel}}}

\affil[3]{\orgdiv{School of Materials Engineering},\orgname{Purdue University},\orgaddress{\city{West Lafayette},\state{Indiana},\postcode{47907}, \country{USA}}}

\affil[4]{\orgdiv{Department of Materials Engineering},\orgname{Indian Institute of Science},\orgaddress{\city{Bengaluru}, \postcode{560012}, \country{India}}}

\affil[5]{\orgdiv{School of Mechanical Engineering},\orgname{Purdue University},\orgaddress{\city{West Lafayette}, \state{Indiana}, \postcode{47907}, \country{USA}}}

\abstract{MXenes have attracted considerable interest for integrated nonlinear photonics owing to their broadband optical response and chemical tunability, yet investigations have focused predominantly on Ti-based compositions, leaving the nonlinear potential of double-transition-metal MXenes largely unexplored. Here we quantify the nonlinear optical response of the double-transition-metal MXene \ce{Mo2TiC2T_x} and demonstrate its functionality as a waveguide-integrated nonlinear activation element for neuromorphic photonics. Z-scan measurements at \SI{800}{\nano\metre} reveal strong saturable absorption in films of two different thicknesses, yielding effective nonlinear absorption coefficients ($\beta_\mathrm{eff}$) between approximately $-2.69\times10^{3}$ and $-0.88\times10^{3}$ \si{\centi\metre\per\giga\watt}, with the nonlinear response decreasing at higher excitation intensities. The extracted nonlinear absorption is approximately one order of magnitude larger than that reported for \ce{Ti3C2T_x} MXene under comparable conditions. We integrate an ultrathin \ce{Mo2TiC2T_x} MXene layer onto a silicon rib waveguide to realize a compact nonlinear optical activation function, which, when implemented in a neural-network emulator, achieves $98.39\%$ classification accuracy on the MNIST benchmark. This work expands the MXene material platform beyond Ti-based compositions and establishes double-transition-metal MXenes as promising candidates for high-performance integrated nonlinear photonic and neuromorphic computing technologies.}


\maketitle

\section{Introduction}
MXenes are a chemically diverse family of two-dimensional (2D) transition-metal carbides, nitrides, and carbonitrides, typically described by the formula $M_{n+1}X_nT_x$, where $M$ is a transition metal, $X$ is carbon or nitrogen, and $T_x$ denotes surface terminations \cite{ref:naguib2011two, ref:anasori20172d, ref:hantanasirisakul2018electronic}. Their electronic and optical properties can be tuned through transition-metal composition and surface chemistry \cite{ref:anasori20172d, ref:hantanasirisakul2018electronic, ref:han2020tailoring, ref:li2024mxenes, ref:zhang2022mxenesphotonics}. In photonics, MXenes exhibit broadband optical absorption and intensity-dependent nonlinear responses, including saturable absorption (SA) \cite{ref:hantanasirisakul2018electronic, ref:jiang2018broadband, ref:zhang2022mxenesphotonics, ref:jin2024thickness}. These responses have enabled broadband optical switching, femtosecond pulse generation, all-optical wavelength conversion, and ultrafast all-optical modulation \cite{ref:jiang2018broadband, ref:jhon2017metallic, ref:song2019nonlinear, ref:sun2025hot}. A continuing goal in integrated photonics is to integrate such nonlinear functionality into compact, chip-scale architectures \cite{ref:bogaerts2020programmable, ref:pelgrin2023hybrid}. In this context, MXenes can be incorporated as thin functional layers on established photonic platforms, with nonlinear responses already demonstrated in silicon and silicon-nitride waveguides \cite{ref:hazan2023mxene, ref:jin2024thickness}.

Despite this progress, nonlinear optical studies of MXenes have focused primarily on Ti-based compositions, particularly \ce{Ti3C2T_x} MXene, for which broadband SA and dependencies on surface termination, film thickness, and excitation conditions have been reported \cite{ref:zhang2025mxenes, ref:jiang2018broadband, ref:li2022switching, ref:jin2024thickness}. Ordered double-transition-metal MXenes extend this compositional space by incorporating chemically distinct transition metals in ordered atomic layers \cite{ref:anasori2015two}. More recently, the related ordered \ce{Mo2Ti2C3T_x} MXene has exhibited ultrafast nonlinear optical responses under visible and infrared femtosecond excitation \cite{ref:stavrou2025emerging}. Nonlinear photonic functionality has also been demonstrated using a \ce{Mo2TiC2}/poly(vinyl alcohol) composite saturable absorber on a side-polished fiber for mode-locked pulse generation \cite{ref:lee2023passively}. Together, these studies highlight the nonlinear photonic potential of Mo-containing ordered MXenes, yet the quantitative femtosecond characterization of nonlinear absorption in continuous \ce{Mo2TiC2T_x} MXene thin films remains unresolved, including the magnitude of the effective nonlinear absorption coefficient ($\beta_{\mathrm{eff}}$) and its dependence on excitation intensity.

Nonlinear optical materials are also being explored as activation elements in photonic neural networks (NNs), where linear operations such as matrix multiplication can be implemented optically, but nonlinear activation remains a central hardware challenge \cite{ref:shen2017deep, ref:williamson2019reprogrammable}. For example, a few-layer \ce{MoTe2} integrated with an optical waveguide has enabled broadband all-optical nonlinear activation \cite{ref:chen2024ultra}. More recently, vertically grown \ce{MoS2} waveguides have demonstrated broadband all-optical nonlinear activation across the telecommunications O- and C-bands \cite{ref:zhang2026integrated}. Within MXenes, \ce{Ti3C2T_x} has been integrated with an optical microfiber to realize broadband ultrafast nonlinear activation \cite{ref:yang2022mxene}. \ce{Ti3C2T_x} nanoflakes deposited on a silicon rib waveguide have also yielded wavelength-dependent nonlinear transfer functions that were evaluated as activation functions in NN emulators \cite{ref:hazan2023mxene}. Together, these studies establish \ce{Ti3C2T_x} as a viable MXene platform for nonlinear optical activation. However, this functionality has yet to be demonstrated using a continuous \ce{Mo2TiC2T_x} ordered double-transition-metal MXene thin film integrated with a silicon waveguide.

Here we establish the ordered double-transition-metal MXene \ce{Mo2TiC2T_x} as a platform for nonlinear photonics. Femtosecond open-aperture (OA) Z-scan measurements at \SI{800}{\nano\meter} reveal pronounced SA in continuous \ce{Mo2TiC2T_x} MXene thin films. The extracted $\beta_{\mathrm{eff}}$ decreases in magnitude with increasing excitation intensity and reaches a maximum of approximately $2.69\times10^{3}$~\si{\centi\meter\per\giga\watt}, nearly one order of magnitude larger than that reported for \ce{Ti3C2T_x} MXene under comparable femtosecond excitation conditions \cite{ref:jin2024thickness}. When incorporated as a continuous thin film on a silicon rib waveguide, \ce{Mo2TiC2T_x} exhibits power- and wavelength-dependent nonlinear transmission, producing saturating transfer functions at representative wavelengths between \SI{1250}{\nano\meter} and \SI{1550}{\nano\meter}. Evaluating these experimentally derived transfer functions as nonlinear activations in a computational NN emulator yielded $98.39\%$ accuracy for the Modified National Institute of Standards and Technology (MNIST) handwritten-digit classification. Together, these results expand the nonlinear photonic materials space of MXenes beyond predominantly studied Ti-based compositions to ordered double-transition-metal chemistry.

\section{Results}\label{sec:results}
\subsection{Material preparation and characterization}

\begin{figure}[htbp]
    \centering
    \includegraphics[width=\textwidth]{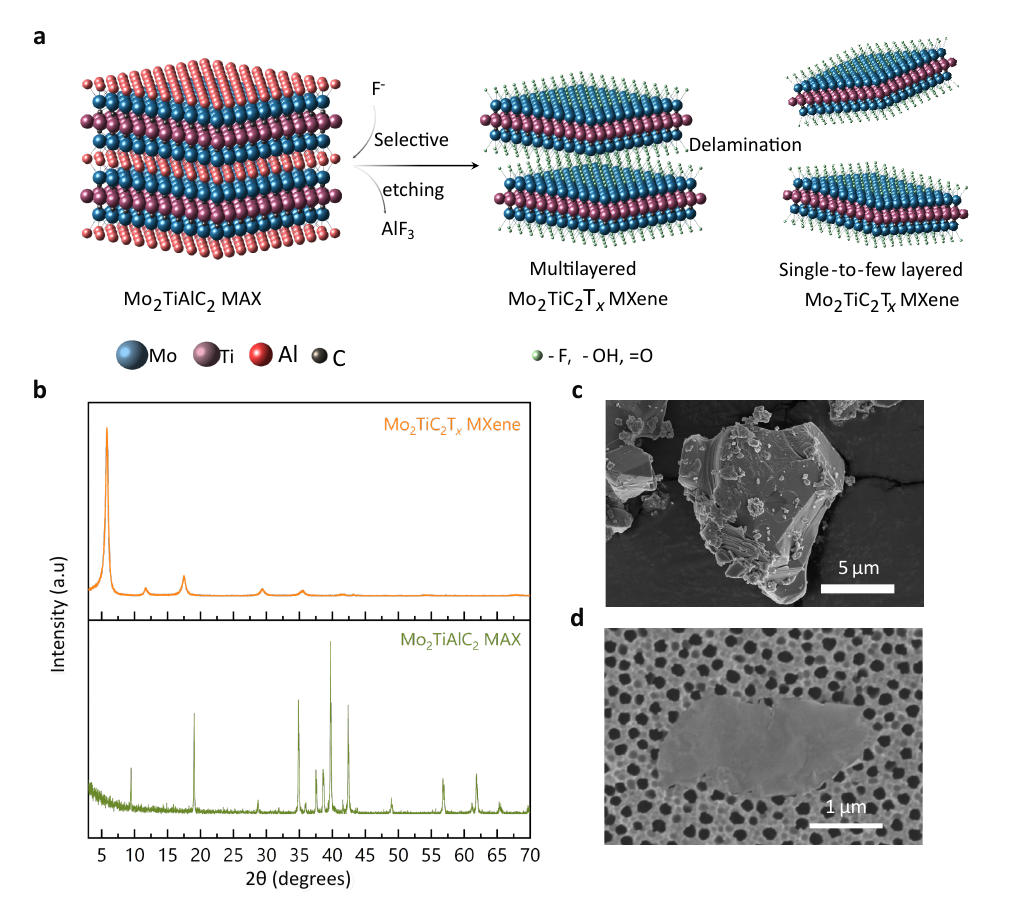}
    \caption{\textbf{Synthesis and characterization of \ce{Mo2TiC2T_x} MXene.} 
    \textbf{a,} Schematic representation of the conversion of \ce{Mo2TiAlC2} MAX to multilayered \ce{Mo2TiC2T_x} MXene by selective etching, followed by delamination to obtain single-to-few-layered \ce{Mo2TiC2T_x} MXene flakes. 
    \textbf{b,} X-ray diffraction patterns of \ce{Mo2TiAlC2} MAX and \ce{Mo2TiC2T_x} MXene. 
    \textbf{c,} FESEM image showing the layered morphology of the \ce{Mo2TiAlC2} MAX precursor. 
    \textbf{d,} FESEM image of a delaminated single-to-few-layered \ce{Mo2TiC2T_x} MXene flake.}
    \label{fig:material}
\end{figure}

\ce{Mo2TiC2T_x} MXene was synthesized from \ce{Mo2TiAlC2} MAX by selective etching followed by delamination, as summarized schematically in Fig.~\ref{fig:material}a (see Methods for synthesis details). The diffraction pattern of \ce{Mo2TiAlC2} MAX exhibited characteristic (00L) peaks for \ce{M3AlC2}, with a c-axis lattice parameter (c-LP) of \SI{18.61}{\angstrom}. Notably, the (002) peaks shifted to lower $2\theta$ angles from approximately $9.6^{\circ}$ to $5.92^{\circ}$ after selective etching of the Al layer and delamination, indicating the formation of \ce{Mo2TiC2T_x} MXene with a c-LP of approximately \SI{30}{\angstrom}, as shown in Fig.~\ref{fig:material}b (see Methods for X-ray diffraction characterization).

Field-emission scanning electron microscopy (FESEM) micrographs of the \ce{Mo2TiAlC2} MAX phase reveal the layered structure of the bulk carbide (Fig.~\ref{fig:material}c). After selective etching and delamination, FESEM images confirm single- to few-layered \ce{Mo2TiC2T_x} MXene flake morphology (Fig.~\ref{fig:material}d). FESEM measurement and sample-preparation details are provided in the Methods.

The delaminated \ce{Mo2TiC2T_x} MXene was then deposited on borosilicate substrates to form continuous thin films for optical characterization and nonlinear measurements. Films prepared with two and six deposition cycles, denoted S1 and S2, respectively, had thicknesses of $\approx$\SI{13}{\nano\meter} and \SI{32}{\nano\meter} and were used for nonlinear optical measurements. Details of thin-film deposition and thickness measurements are provided in the Methods section. Additional physical and optical characterization, including ultraviolet–visible transmission spectra, is provided in Supplementary Note~1.

\subsection{Femtosecond saturable absorption}

\begin{figure}[htbp]
    \centering
    \includegraphics[width=\textwidth]{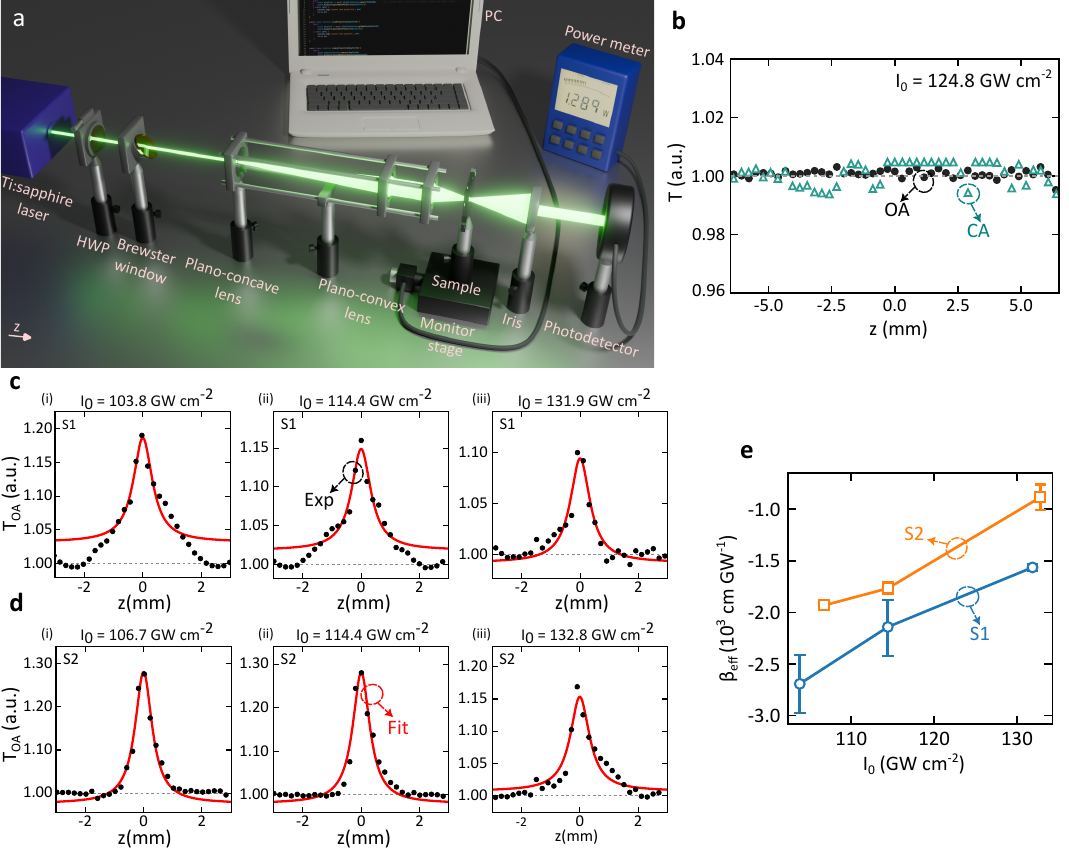}
    \caption{\textbf{Femtosecond saturable absorption of \ce{Mo2TiC2T_x} MXene thin films.}
    \textbf{a,} Experimental configuration for the Z-scan measurements at \SI{800}{\nano\meter} using \SI{140}{\femto\second} pulses at a repetition rate of \SI{80}{\mega\hertz}.
    \textbf{b,} Normalized OA and CA transmittance of the uncoated borosilicate substrate at an incident peak intensity of $I_0\approx$\SI{124.8}{\giga\watt\per\square\centi\meter}.
    \textbf{c,d,} Normalized OA transmittance of the \ce{Mo2TiC2T_x} MXene films S1 and S2, respectively, measured at increasing incident peak intensities: \textbf{c,} (i) 103.8, (ii) 114.4 and (iii) \SI{131.9}{\giga\watt\per\square\centi\meter}; \textbf{d,} (i) 106.7, (ii) 114.4 and (iii) \SI{132.8}{\giga\watt\per\square\centi\meter}.
    Black circles represent the experimental measurements, red curves show the OA Z-scan fits, and the grey dashed line denotes unity normalized transmittance.
    \textbf{e,} Extracted $\beta_{\mathrm{eff}}$ as a function of $I_0$ for S1 and S2.
    Points show the mean values obtained from $n=3$ spatial positions across the coated region; error bars denote one standard deviation, and lines are guides to the eye. HWP, half-wave plate.}
    \label{fig:zscan}
\end{figure}

To quantify the femtosecond nonlinear absorption of continuous \ce{Mo2TiC2T_x} MXene thin films, we performed OA Z-scan measurements at \SI{800}{\nano\meter} using \SI{140}{\femto\second} pulses at a repetition rate of \SI{80}{\mega\hertz}, with the experimental configuration shown in Fig.~\ref{fig:zscan}a (see Methods for experimental details). An uncoated borosilicate substrate was measured under the same excitation conditions to assess the substrate contribution. At an incident peak intensity of $I_0\approx$\SI{124.8}{\giga\watt\per\square\centi\meter}, both the OA and closed-aperture (CA) traces remained nearly flat (Fig.~\ref{fig:zscan}b), indicating no appreciable nonlinear optical response from the substrate under these conditions. Independent validation of the Z-scan configuration using reference samples is provided in Supplementary Note~2.

In contrast to the substrate, both \ce{Mo2TiC2T_x} MXene films exhibited pronounced transmission maxima near the focal position across the investigated excitation intensities (Fig.~\ref{fig:zscan}c(i--iii), d(i--iii)). Because the optical intensity is highest near focus, the corresponding increase in normalized transmittance indicates intensity-induced bleaching of the absorptive response, establishing the characteristic OA signature of SA. The OA Z-scan model captured the principal transmission response of both films, although small deviations from the ideal symmetric line shape remained in some measurements.

The extracted $\beta_{\mathrm{eff}}$ values as a function of $I_0$ are shown in Fig.~\ref{fig:zscan}e. Negative $\beta_{\mathrm{eff}}$ values were obtained for both \ce{Mo2TiC2T_x} MXene films throughout the investigated intensity range, consistent with their SA response. For S1, $\beta_{\mathrm{eff}}$ changed from $-2.69\times10^{3}$ to $-1.56\times10^{3}$~\si{\centi\meter\per\giga\watt} as $I_0$ increased from $103.79$ to $131.87$~\si{\giga\watt\per\square\centi\meter}, whereas for S2 it changed from $-1.93\times10^{3}$ to $-0.88\times10^{3}$~\si{\centi\meter\per\giga\watt} as $I_0$ increased from $106.73$ to $132.82$~\si{\giga\watt\per\square\centi\meter}. In both films, the magnitude of $\beta_{\mathrm{eff}}$ decreased overall with increasing $I_0$, consistent with progressive saturation of the nonlinear absorption response. At comparable $I_0$ values, S1 exhibited a larger magnitude of $\beta_{\mathrm{eff}}$ than S2.

\subsection{Waveguide-integrated nonlinear optical activation}

\begin{figure}[htbp]
    \centering
    \includegraphics[width=\textwidth]{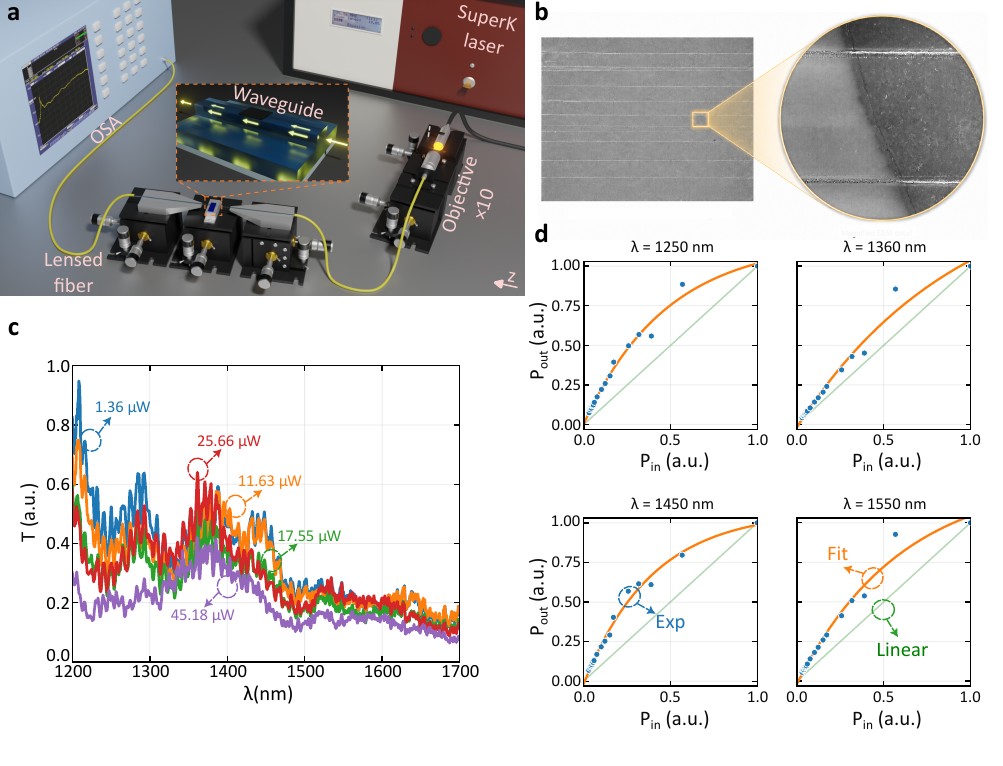}
    \caption{\textbf{Waveguide-integrated nonlinear optical response of \ce{Mo2TiC2T_x} MXene.}
    \textbf{a,} Experimental configuration used for the waveguide-transmission measurements.
    \textbf{b,} SEM image of the continuous \ce{Mo2TiC2T_x}MXene-coated region of the silicon rib waveguide.
    \textbf{c,} Normalized transmission spectra, $T$, of the \ce{Mo2TiC2T_x} MXene-coated waveguide relative to the uncoated reference waveguide at different calibrated input powers.
    \textbf{d,} Normalized transfer functions extracted at \SI{1250}{\nano\meter}, \SI{1360}{\nano\meter}, \SI{1450}{\nano\meter} and \SI{1550}{\nano\meter}.
    Blue symbols represent the experimentally derived responses, orange curves show the fit, and the light-green diagonal denotes linear scaling. OSA, optical spectrum analyzer.}
    \label{fig:waveguide}
\end{figure}

A silicon rib waveguide coated with a continuous \ce{Mo2TiC2T_x} MXene thin film exhibited transmission that varied with both input power and wavelength. The waveguide was \SI{400}{\nano\meter} thick, \SI{10}{\micro\meter} wide, and \SI{7}{\milli\meter} long, with an approximately \SI{13}{\nano\meter}-thick \ce{Mo2TiC2T_x} MXene film extending over approximately \SI{500}{\micro\meter} of its length (Fig.~\ref{fig:waveguide}b). SEM confirmed that the film remained continuous across the defined coated region (see Methods for waveguide fabrication and SEM characterization). The transmission response was measured using the configuration shown in Fig.~\ref{fig:waveguide}a at calibrated waveguide input powers ranging from \SI{1.36}{\micro\watt} to \SI{45.18}{\micro\watt} (see Methods).

The transmission spectra changed systematically with input power across the measured near-infrared range (Fig.~\ref{fig:waveguide}c). Relative to the uncoated reference waveguide, the \ce{Mo2TiC2T_x} MXene-coated waveguide exhibited additional wavelength-dependent attenuation, and transmission generally decreased as input power increased. The complete transmission spectra are provided in Supplementary Note~3. The guided optical mode overlaps with the \ce{Mo2TiC2T_x} MXene film through its evanescent field at the upper waveguide interface, providing an interaction pathway through which the film modifies the transmitted signal over the coated section. The extracted transfer functions at \SI{1250}{\nano\meter}, \SI{1360}{\nano\meter}, \SI{1450}{\nano\meter}, and \SI{1550}{\nano\meter} were monotonic and concave down, departing systematically from linear scaling (Fig.~\ref{fig:waveguide}d). This sublinear response produced saturating transfer characteristics suitable for nonlinear activation. Each transfer function was represented by a saturating exponential fit for the following computational implementation.

\begin{figure}[htbp]
    \centering
    \includegraphics[width=\textwidth]{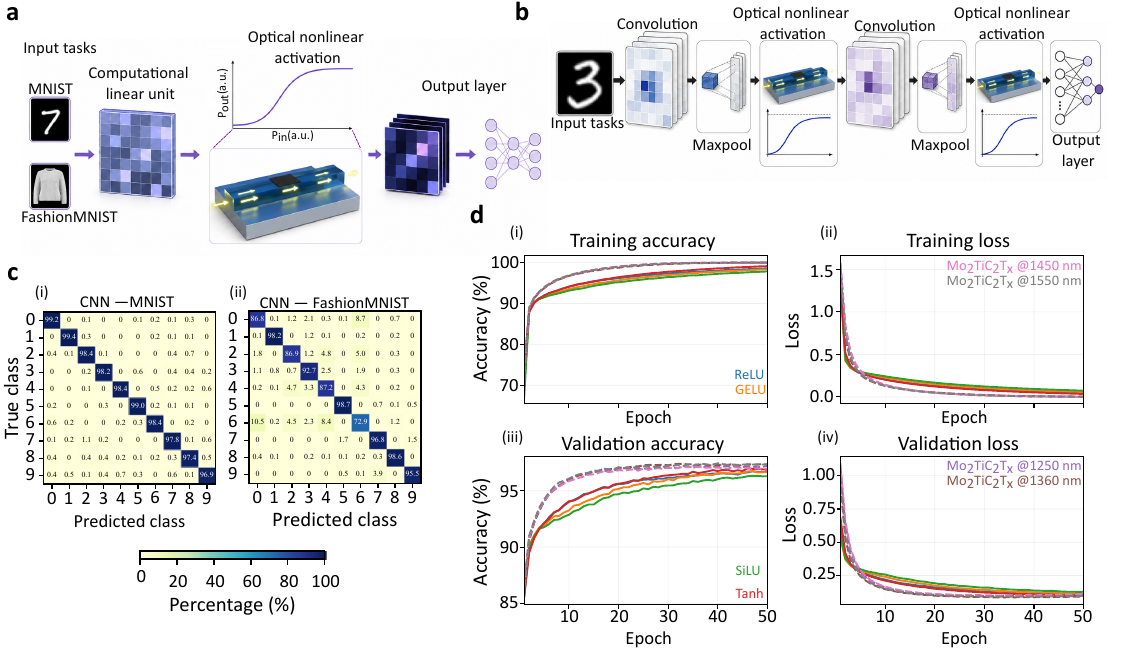}
    \caption{\textbf{Neural-network evaluation of experimentally derived \ce{Mo2TiC2T_x} MXene waveguide activations.}
    \textbf{a,} Experimental--computational workflow in which the linear NN operations are performed computationally, and the fitted wavelength-dependent waveguide response provides the nonlinear activation.
    \textbf{b,} Simplified CNN workflow illustrating the use of the experimentally derived waveguide activation at the nonlinear stages.
    \textbf{c,} Test-set confusion matrices for CNN models using the \SI{1550}{\nano\meter} waveguide response on MNIST and Fashion-MNIST.
    Values are normalized by true class and reported as percentages.
    \textbf{d,} Representative training and validation accuracy and loss histories comparing the MXene-waveguide activations with ReLU, Tanh, GELU, and SiLU.}
    \label{fig:nn_emulator}
\end{figure}

To evaluate these experimentally derived responses as nonlinear activations, the fitted waveguide transfer functions were implemented in a computational NN emulator, while linear operations were computed numerically (Fig.~\ref{fig:nn_emulator}a). The four wavelength-dependent activation functions were evaluated in both a feedforward neural network (FFNN) and a convolutional neural network (CNN), with the CNN workflow illustrated schematically in Fig.~\ref{fig:nn_emulator}b. The networks were evaluated on MNIST and Fashion-MNIST and compared with otherwise equivalent models using ReLU, Tanh, GELU, and SiLU activation functions. Details of the network architectures, training procedure, and implementation of the MXene-waveguide activation functions are provided in the Methods and Supplementary Note~4.

The \ce{Mo2TiC2T_x} MXene activation functions were evaluated across both network architectures and datasets. For MNIST, the CNN achieved $98.39\%$ accuracy with the \SI{1450}{\nano\meter} response, compared with $98.08\%$ for Tanh, the best-performing conventional activation in this architecture. The FFNN achieved $97.47\%$ with the \SI{1550}{\nano\meter} response, compared with $97.36\%$ for Tanh. For Fashion-MNIST, the highest accuracies were $88.53\%$ in the FFNN with the \SI{1360}{\nano\meter} response and $91.90\%$ in the CNN with the \SI{1450}{\nano\meter} response, compared with $87.70\%$ for Tanh and $91.29\%$ for ReLU, respectively. Representative test-set confusion matrices for the CNN with the \SI{1550}{\nano\meter} response are shown in Fig.~\ref{fig:nn_emulator}c, while representative training and validation histories are shown in Fig.~\ref{fig:nn_emulator}d. Additional training and validation histories for the remaining architecture--dataset combinations are provided in Supplementary Note~4. The wavelength-dependent waveguide responses therefore provide a selectable family of experimentally derived nonlinear activations rather than defining a single universally optimal operating wavelength.

\section{Discussion}\label{sec:discussion}

\begin{figure}[htbp]
    \centering
    \includegraphics[width=\textwidth]{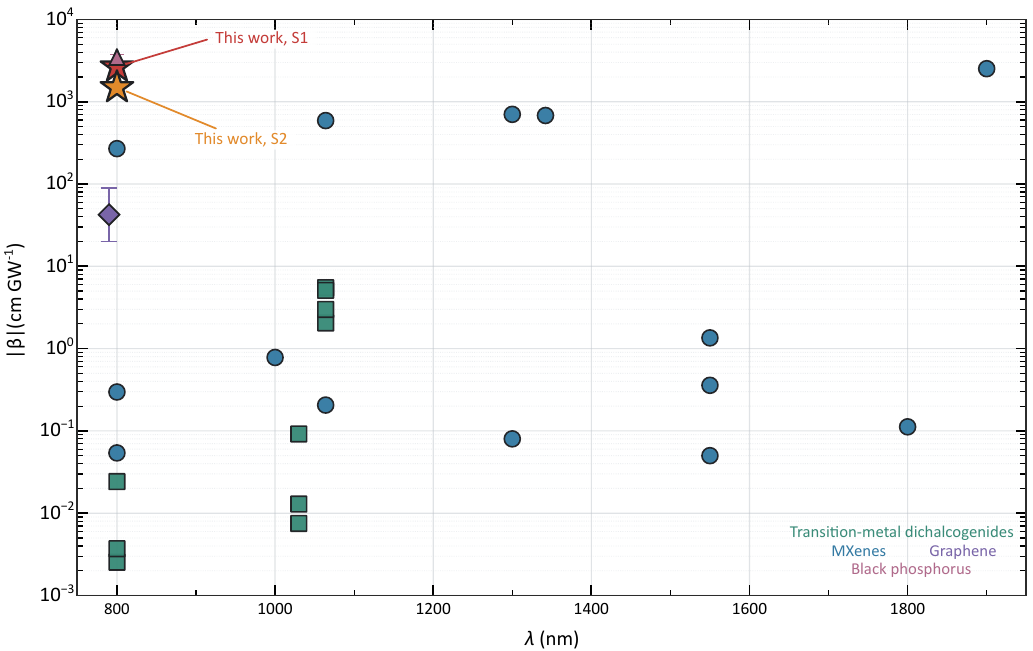}
    \caption{\textbf{Nonlinear absorption in representative two-dimensional materials.}
    Magnitude of the nonlinear absorption coefficient, $|\beta|$, as a function of excitation wavelength for the \ce{Mo2TiC2T_x} MXene films investigated in this work and representative 2D materials reported in the literature~\cite{ref:jin2024thickness, ref:zhang2021third, ref:gao2022optical, ref:wang20232d, ref:kumar2009femtosecond, ref:wang2014broadband, ref:bikorimana2016nonlinear}. The \ce{Mo2TiC2T_x} values correspond to the largest magnitude of $\beta_{\mathrm{eff}}$ for each film. Literature points represent negative nonlinear absorption coefficients measured under the experimental conditions reported in the corresponding studies.}
    \label{fig:beta_comparison}
\end{figure}

Ordered \ce{Mo2TiC2T_x} MXene exhibits a strong nonlinear absorption response relative to the closest \ce{Ti3C2T_x} MXene thin-film benchmark, measured under closely comparable femtosecond excitation conditions. An approximately \SI{90}{\nano\meter} \ce{Ti3C2T_x} MXene film exhibited $\beta_{\mathrm{eff}}\approx-2.69\times10^{2}$~\si{\centi\meter\per\giga\watt} near \SI{800}{\nano\meter} under \SI{140}{\femto\second}, \SI{80}{\mega\hertz} excitation, whereas \ce{Mo2TiC2T_x} MXene reaches $-2.69\times10^{3}$~\si{\centi\meter\per\giga\watt}, corresponding to an order-of-magnitude increase in $\beta_{\mathrm{eff}}$~\cite{ref:jin2024thickness}. Placing this response within the broader landscape of 2D materials, Fig.~\ref{fig:beta_comparison} compares the present \ce{Mo2TiC2T_x} films with representative MXenes, graphene, black phosphorus, and transition-metal dichalcogenides reported in the literature~\cite{ref:zhang2021third, ref:gao2022optical, ref:wang20232d, ref:kumar2009femtosecond, ref:wang2014broadband, ref:bikorimana2016nonlinear}. Although the reported coefficients were obtained under different excitation conditions, film thicknesses, and sample configurations, the comparison places the \ce{Mo2TiC2T_x} response among the larger nonlinear absorption magnitudes reported for these representative systems. In the present samples, the larger magnitude of $\beta_{\mathrm{eff}}$ for the thinner \ce{Mo2TiC2T_x} MXene film further indicates that the nonlinear response does not simply increase with the deposited material thickness. Together with previous reports of thickness-dependent nonlinear absorption in \ce{Ti3C2T_x} MXene~\cite{ref:jin2024thickness, ref:ma2022passively, ref:chinnapaiyan2026third} and the sensitivity of $\beta_{\mathrm{eff}}$ to surface termination and surface chemical modification~\cite{ref:li2022switching, ref:zhao2024enhanced, ref:shan2024n}, these observations indicate that film thickness and surface chemistry can influence nonlinear absorption in MXene thin films.

Integrating a continuous, ordered \ce{Mo2TiC2T_x} MXene film with a silicon rib waveguide yields sublinear guided-wave transfer functions, demonstrating nonlinear optical functionality in an integrated photonic geometry. Hazan \textit{et al.} previously demonstrated nonlinear activation using \ce{Ti3C2T_x} MXene nanoflakes on a silicon rib waveguide over an approximately \SI{7}{\milli\meter} interaction length, whereas the continuous \ce{Mo2TiC2T_x} film used here produces the nonlinear transfer response within an approximately \SI{500}{\micro\meter} coated region, reducing the longitudinal footprint of the active section by approximately fourteen-fold~\cite{ref:hazan2023mxene}. Other 2D-material nonlinear activators have employed butt-coupled material--waveguide geometries or resonant photonic structures to enhance light-matter interaction~\cite{ref:chen2024ultra, ref:liu2026femto}; in contrast, the present response is obtained through evanescent interaction with the MXene film on a straight, non-resonant silicon rib waveguide, without electrical or thermal control. The operating wavelength further provides access to distinct nonlinear mappings from the same physical element. Although wavelength-dependent nonlinear activation has previously been demonstrated in \ce{Ti3C2T_x} MXene devices~\cite{ref:hazan2023mxene, ref:yang2022mxene}, here it is realized using a continuous, ordered, double-transition-metal \ce{Mo2TiC2T_x} film integrated on silicon. When implemented in the computational NN emulator, the experimentally derived transfer functions remained effective as nonlinear activations, reaching classification accuracies of up to $98.39\%$, with the preferred wavelength varying with network architecture and dataset.

More broadly, ordered double-transition-metal MXenes expand the compositional space for nonlinear photonics beyond the predominantly studied Ti-based systems. The established sensitivity of MXene optical properties to transition-metal composition and surface termination suggests that these parameters may offer routes to control their nonlinear optical response~\cite{ref:han2020tailoring, ref:li2022switching}. Systematic studies across ordered MXene compositions could therefore clarify how composition, surface chemistry, and film structure govern nonlinear absorption and guided-wave optical responses, providing a basis for tailoring MXene materials to specific nonlinear photonic functions.

\section{Methods}\label{sec:methods}

\subsection{Synthesis and preparation of \texorpdfstring{\ce{Mo2TiC2T_x}}{Mo2TiC2Tx} MXene}

\subsubsection{MAX-phase synthesis}
\ce{Mo2TiAlC2} MAX was synthesized using pressureless sintering in an argon atmosphere. Molybdenum, titanium, aluminum, and calcined coke powder as the carbon source were mixed in a molar ratio of $2.2:0.8:1.1:2$ and jar-milled for \SI{18}{\hour} at 60~rpm using zirconia balls. The powders were then sintered in a tube furnace (Carbolite Gero, \SI{1700}{\degreeCelsius} model) at \SI{1600}{\degreeCelsius} for \SI{4}{\hour} under a constant argon flow. The sintered MAX block was subsequently milled to yield fine MAX powder using a \SI{71}{\micro\meter} sieve before MXene synthesis.

\subsubsection{MXene synthesis}
To synthesize \ce{Mo2TiC2T_x} MXene, \SI{1}{\gram} of \ce{Mo2TiAlC2} MAX phase was mixed with \SI{10}{\milli\liter} of hydrofluoric acid (HF, 49--51~wt\%, Fisher Scientific) as an etchant in a high-density polyethylene bottle and stirred at 300~rpm for \SI{96}{\hour} at \SI{55}{\degreeCelsius}. The etched multilayered \ce{Mo2TiC2T_x} MXene flakes were washed with deionized water through repeated centrifugation at 3234~RCF (4-5 cycles with approximately \SI{300}{\milli\liter} of deionized water) until the supernatant reached pH~$\sim7$.

To delaminate, the etched multilayer \ce{Mo2TiC2T_x} MXene sediment was added to \SI{5}{\milli\liter} of tetramethylammonium hydroxide (TMAOH) solution (25~wt\% stock, Fisher Scientific) in \SI{20}{\milli\liter} of deionized water per gram of starting \ce{Mo2TiAlC2} MAX. The mixture was stirred at 300~rpm for \SI{4}{\hour} at \SI{55}{\degreeCelsius}. After delamination, the TMA$^{+}$ intercalated \ce{Mo2TiC2T_x} MXene solution was washed to neutral pH by repeated centrifugation at 21900~RCF (4 cycles with approximately \SI{300}{\milli\liter} of deionized water). Thereafter, the final mixture of \ce{Mo2TiC2T_x} MXene was re-dispersed in \SI{20}{\milli\liter} of deionized water and vortexed for \SI{15}{\minute}. The suspension was centrifuged at 3000~RCF for \SI{30}{\minute} to collect single-to-few-layered \ce{Mo2TiC2T_x} MXene flakes. The final \ce{Mo2TiC2T_x} MXene suspension was collected and stored at \SI{-20}{\degreeCelsius} until use.

\subsubsection{Thin-film deposition}
\ce{Mo2TiC2T_x} MXene deposition was performed using the method demonstrated in our previous study~\cite{ref:favelukis2025without}. Borosilicate dies (\SI{1}{\square\centi\meter}) were treated with a 1:3 \ce{H2O2}/\ce{H2SO4} piranha solution (\ce{H2SO4}, 98\%, SDFCL; \ce{H2O2}, 35\%, Thermo Fisher Scientific) to increase their hydrophilicity. The dies were spin-coated multiple times with the MXene solution at 2000~rpm, followed by spin-cleaning with 0.5~M HCl (32\%, Bio-Lab) at 2000~rpm. 

\subsection{X-ray diffraction characterization}
X-ray diffraction (XRD) patterns of the as-synthesized \ce{Mo2TiAlC2} MAX phase and \ce{Mo2TiC2T_x} MXene were obtained using an XRDynamic 500 diffractometer (Anton Paar) with Cu K$\alpha$ radiation ($\lambda=1.5406$~\AA). The samples were mounted on the fixed sample stage and scanned from $3^\circ$ to $70^\circ$ with a step size of $0.01^\circ$ and a time per step of \SI{20}{\second}.

\subsection{Field-emission scanning electron microscopy}
Field-emission scanning electron microscopy (FESEM) was performed using a JEOL JSM-7800f FESEM with a low-beam detector at acceleration voltages of \SI{5}{\kilo\volt} and \SI{15}{\kilo\volt} to study the surface morphology and flake size. The \ce{Mo2TiC2T_x} MXene solution concentration was maintained below \SI{0.1}{\milli\gram\per\milli\liter}, and the solution was loaded onto an anodic disc, followed by vacuum drying for \SI{2}{\hour}. The samples were gold-sputtered to reduce charging and improve the sharpness of the SEM images.

\subsection{Thin-film thickness characterization}
The thickness and surface topography of the \ce{Mo2TiC2T_x} MXene thin films were measured using a stylus profilometer (Dektak 8, Veeco). The measured film thicknesses were $\approx 13$~\si{\nano\meter} and $32$~\si{\nano\meter} for S1 and S2, respectively. Additional profilometry and atomic force microscopy characterization is provided in Supplementary Note~1.

\subsection{Spectroscopic ellipsometry}
The optical properties of the \ce{Mo2TiC2T_x} MXene thin films were characterized by spectroscopic ellipsometry over the wavelength range of \SIrange{250}{1700}{\nano\meter}. The measured ellipsometric parameters were fitted using a multilayer optical model comprising the borosilicate substrate and the \ce{Mo2TiC2T_x} MXene film. The extracted optical constants and details of the ellipsometric fitting are provided in Supplementary Note~1.

\subsection{Silicon rib-waveguide fabrication} 
The silicon rib waveguides were fabricated on a silicon-on-insulator wafer using a previously established fabrication procedure~\cite{ref:katiyi2018si}. The wafer consisted of a silicon handle substrate, a \SI{2}{\micro\meter}-thick buried \ce{SiO2} layer and a \SI{2}{\micro\meter}-thick silicon device layer. Poly (methyl methacrylate) resist (PMMA 950k) was deposited on the wafer and patterned to define the waveguide structures. Following resist development, a \SI{250}{\nano\meter}-thick aluminum layer was deposited by electron-beam evaporation and used as a hard mask during silicon etching. Lift-off was performed by immersing the chip in acetone for \SI{4}{\hour}, then rinsing it with isopropanol and drying it. The silicon device layer was subsequently dry-etched using a plasma containing \ce{SF6}, Ar, and \ce{O2}, leaving a \SI{400}{\nano\meter}-thick silicon rib with nominally vertical sidewalls. The residual aluminum hard mask was removed using 400K developer.

\subsection{Scanning electron microscopy}
The surface morphology of the uncoated reference waveguides and the \ce{Mo2TiC2T_x} MXene-coated rib waveguides was examined using a high-resolution scanning electron microscope (Verios 460L, FEI). The SEM images were used to assess the waveguide surface, the morphology of the deposited \ce{Mo2TiC2T_x} MXene layer, and the spatial extent of the coated section.

\subsection{Open-aperture Z-scan measurements}
The nonlinear absorption of the \ce{Mo2TiC2T_x} MXene thin films was characterized using the OA Z-scan technique introduced by Sheik-Bahae \textit{et al.}~\cite{ref:sheik1990sensitive}. Measurements were performed with a Ti:sapphire femtosecond laser (Chameleon Ultra II, Coherent) operating at a central wavelength of \SI{800}{\nano\meter}, with a pulse duration of \SI{140}{\femto\second} and a repetition rate of \SI{80}{\mega\hertz}. Incident optical power was controlled with a zero-order achromatic half-wave plate (AHWP05M-980, Thorlabs) mounted in a precision rotation mount (CRM1PT, Thorlabs), followed by an ultraviolet fused-silica Brewster window (BW0801, Thorlabs) held in a Brewster-window mount (BW16M, Thorlabs). Before focusing, the beam was expanded with a $6\times$ telescope composed of a plano-concave lens with a focal length of \SI{-25}{\milli\meter} (LC1054-B-ML, Thorlabs) and a plano-convex lens with a focal length of \SI{150}{\milli\meter} (LA1433-B-ML, Thorlabs). The expanded beam was then focused onto the sample with a \SI{150}{\milli\meter} focal-length lens (LA1433-B-ML, Thorlabs).

Each \ce{Mo2TiC2T_x} MXene sample was mounted normal to the incident beam on a motorized linear translation stage with a \SI{100}{\milli\meter} travel range (NRT100, Thorlabs). The stage was driven by a K-Cube stepper-motor controller (KST201, Thorlabs) and translated the sample along the beam-propagation direction through the focal region. A color CMOS camera (CS165CU, Thorlabs) assisted with optical alignment and observation of the focal region. For OA measurements, the full transmitted beam was collected without an aperture and measured with a Coherent photodetector connected to a Coherent optical power meter. The calculated $1/e^2$ beam-waist radius at the focal plane was approximately \SI{10.61}{\micro\meter}, corresponding to a Rayleigh length of $z_{\mathrm{R}}\approx$\SI{0.442}{\milli\meter}. The on-axis peak intensity, $I_0$, was calculated from the measured average incident power using the laser repetition rate, pulse duration, and the calculated beam-waist radius. 

The normalized OA transmittance was fitted using the thin-sample Z-scan model~\cite{ref:sheik1990sensitive},
\begin{equation}
    T_{\mathrm{OA}}(z)=\sum_{m=0}^{\infty}\frac{1}{(m+1)^{3/2}}\left[-\frac{\beta_{\mathrm{eff}}I_0L_{\mathrm{eff}}}{1+\left[(z-z_c)/z_{\mathrm{R}}\right]^2}\right]^m
    \label{eq:oa_zscan}
\end{equation}
where $T_{\mathrm{OA}}(z)$ is the normalized OA transmittance, $L_{\mathrm{eff}}$ is the effective interaction length, and $z_c$ is the focal-position offset. The effective interaction length was calculated from the physical film thickness measured by stylus profilometry and the linear absorption coefficient derived from the extinction coefficient measured by spectroscopic ellipsometry. The film thicknesses and optical constants used in this calculation were determined as described above, with additional characterization provided in Supplementary Note~1.

At each incident intensity, OA measurements were acquired at three spatially distinct positions across the coated region of each film. The transmitted-power traces were normalized to the far-field transmission and fitted independently using Eq.~\ref{eq:oa_zscan}. The values of $I_0$, $z_{\mathrm{R}}$ and $L_{\mathrm{eff}}$ were fixed during fitting, whereas $\beta_{\mathrm{eff}}$ and $z_c$ were treated as fitting parameters. The reported $\beta_{\mathrm{eff}}$ values represent the mean of the three independently fitted measurements at each incident intensity, and the error bars denote one standard deviation.

\subsection{Waveguide nonlinear-transmission measurements}

Broadband unpolarized light from a SuperK Extreme supercontinuum source (NKT Photonics) was launched into the fiber path via a $\times10$ microscope objective (Olympus) with a numerical aperture of 0.25 and coupled to the waveguide input facet via a lensed single-mode fiber (SMF). The input and output were positioned at the corresponding facets of the silicon rib waveguide using manually controlled three-axis fiber-positioning stages (Thorlabs). Fiber-to-waveguide alignment at both facets was monitored with a Pixelink camera. At the same time, the lateral, vertical, and axial positions of the fibers were adjusted to maximize the transmitted optical signal. 

Light emerging from the waveguide output facet was collected via the output SMF and directed to an optical spectrum analyzer (AQ6370D, Yokogawa). Transmission spectra were recorded over the wavelength range of \SIrange{1000}{1700}{\nano\meter}, with a wavelength accuracy of $\pm$\SI{0.01}{\nano\meter} and a minimum wavelength resolution of \SI{0.02}{\nano\meter}. The incident optical power was varied by adjusting the repetition rate of the supercontinuum source. Before waveguide measurements, the optical power for each repetition-rate setting was calibrated by connecting the source directly to the optical spectrum analyzer. Transmission spectra of the uncoated reference waveguide and the \ce{Mo2TiC2T_x} MXene-coated waveguide were then acquired under identical source conditions at each input-power level.

\subsection{Neural-network emulation}

The NN models were implemented in Python using the PyTorch framework~\cite{ref:paszke2019pytorch}. All weighted-summation, bias-addition, convolution, forward-propagation, and backpropagation operations were computed, whereas the nonlinear activation functions were derived from the experimentally measured transfer functions of the integrated \ce{Mo2TiC2T_x} MXene-waveguide. The transfer functions measured at \SI{1250}{\nano\meter}, \SI{1360}{\nano\meter}, \SI{1450}{\nano\meter}, and \SI{1550}{\nano\meter} were fitted independently with saturating exponential functions and implemented as fixed nonlinear activations. A separate network was trained for each wavelength-dependent activation function. For comparison, otherwise identical models were trained using ReLU, Tanh, GELU, and SiLU. For signed computational inputs, the experimentally derived response was extended as an even function, as detailed in Supplementary Note~4.

Two NN architectures were considered. The FFNN received each $28\times28$-pixel grayscale image as a flattened 784-element input vector and comprised two fully connected hidden layers with 100 neurons each, followed by a ten-neuron output layer. The selected activation function was applied after each hidden layer. The CNN comprised three convolutional layers with 32, 64, and 128 filters, respectively. Each convolution used a $3\times3$ kernel, a stride of one, and one-pixel zero padding. The selected activation function was applied after each convolutional layer, and $2\times2$ max pooling with a stride of two was applied after the first and second convolutional layers. The final feature map was flattened and connected to a ten-neuron classification layer. A softmax operation was applied at the output of both architectures to obtain the class probabilities.

For both MNIST~\cite{ref:lecun1998gradient} and Fashion-MNIST~\cite{ref:xiao2017fashion}, the official $60{,}000$-image training set was randomly split into $48{,}000$ training images and $12{,}000$ validation images, corresponding to an 80:20 split. The official $10{,}000$-image test set was excluded from model fitting and hyperparameter selection and used only for final evaluation.

All models were trained for up to 50 epochs with a batch size of 128. The AdamW optimizer~\cite{ref:loshchilov2017decoupled} was used with a weight-decay coefficient of $10^{-4}$. The learning rate was set to $10^{-4}$ for the FFNN models. For the CNN models, learning rates of $10^{-5}$ and $10^{-4}$ were used for MNIST and Fashion-MNIST, respectively. Training minimized the multiclass cross-entropy loss, with trainable network parameters updated by backpropagation~\cite{ref:goodfellow2016deep}. Additional implementation details and the complete training and validation histories are provided in Supplementary Note~4.

\backmatter

\bmhead{Supplementary information}
Supplementary information is available for this paper.

\bmhead{Acknowledgements}
The research was supported by the Bilateral Science Foundation (BSF), Grant No. 2022014.

\section*{Declarations}

\bmhead{Competing interests}
The authors declare no competing interests.

\bmhead{Data availability}
The data supporting this study are available from the corresponding author upon reasonable request.

\bmhead{Code availability}
The code used in this study is available from the corresponding author upon reasonable request.

\bmhead{Author contributions}

Shadad Watad contributed to the conceptualization, methodology, investigation, and formal analysis of the study. Aviad Katiyi contributed to the methodology, investigation, and visualization, including assistance with the experimental configurations and preparation of figures. Bar Favelukis, Anupma Thakur, and Muhammad Sharif Uddin contributed to the synthesis and preparation of the MXene material. Maxim Sokol and Babak Anasori supervised the material synthesis and preparation. Alina Karabchevsky supervised the study and acquired funding. All authors contributed to writing the original draft and to reviewing and editing the manuscript.

\bibliography{sn-bibliography}
\end{document}